\documentclass[twocolumn, pra, supplementary, superscriptaddress]{revtex4-1} 
\usepackage{amsfonts}
\usepackage{amssymb}
\usepackage{amsmath}
\usepackage{epsfig}
\usepackage{color}
\usepackage{graphicx} 
\usepackage{bbold}
\usepackage{psfrag}
\usepackage{mathcomp}
\usepackage{subfigure}
\usepackage{verbatim}
\usepackage{float}
\usepackage{pifont}
\usepackage{physics}
\usepackage{hyperref}
\usepackage{float}
\hypersetup{
	colorlinks=true,
	linkcolor=blue,
	citecolor=blue,
	urlcolor=blue
}
\makeatletter

\newcommand{\Rmnum}[1]{\expandafter\@slowromancap\romannumeral #1@}
\makeatother

\begin{document}
	\preprint{APS/123-QED}
	
	\title{ Observation of topological switch between Weyl semimetal and third-order topological insulator phases}

	\author{Yu-Hong Han}
	\altaffiliation{These authors contributed equally to this work.} 
	\affiliation{State Key Laboratory of Quantum Optics Technologies and Devices, Institute
		of Laser Spectroscopy, Shanxi University, Taiyuan, Shanxi 030006, China}
	
	\author{Yi Li}
	\altaffiliation{These authors contributed equally to this work.}
	\affiliation{State Key Laboratory of Quantum Optics Technologies and Devices, Institute
		of Laser Spectroscopy, Shanxi University, Taiyuan, Shanxi 030006, China}
	
	\author{Feng Mei}
	\email{meifeng@sxu.edu.cn}
	\affiliation{State Key Laboratory of Quantum Optics Technologies and Devices, Institute
		of Laser Spectroscopy, Shanxi University, Taiyuan, Shanxi 030006, China}
	\affiliation{Collaborative Innovation Center of Extreme Optics, Shanxi University, Taiyuan, Shanxi 030006, China}
	
	\author{Liantuan Xiao}
	\affiliation{State Key Laboratory of Quantum Optics Technologies and Devices, Institute
		of Laser Spectroscopy, Shanxi University, Taiyuan, Shanxi 030006, China}
	\affiliation{Collaborative Innovation Center of Extreme Optics, Shanxi
		University, Taiyuan, Shanxi 030006, China}
	
	\author{Suotang Jia}
	\affiliation{State Key Laboratory of Quantum Optics Technologies and Devices, Institute
		of Laser Spectroscopy, Shanxi University, Taiyuan, Shanxi 030006, China}
	\affiliation{Collaborative Innovation Center of Extreme Optics, Shanxi
		University, Taiyuan, Shanxi 030006, China}

\begin{abstract}
Weyl semimetals and higher-order topological insulators represent two fundamental yet distinct classes of topological matter. While both have been extensively studied in classical-wave systems, their coexistence and controllable transition within a single platform remain largely unexplored. Meanwhile, implementing three-dimensional spin-orbit couplings, which is crucial for realizing a broad class of higher-dimensional topological phases, continues to pose significant experimental challenges. Here, we experimentally realize three-dimensional spin-orbit couplings and demonstrate that tuning the dimerized spin-orbit coupling strength enables both the coexistence of and a controllable switch between Weyl semimetal and third-order topological insulator phases. By engineering a three-dimensional circuit metamaterial, we synthesize the required spin-orbit interactions and observe hallmark signatures of both phases: frequency spectroscopy reveals the Fermi arcs, while local density of states measurements identify the topological corner modes. Interestingly, the corner mode degeneracy doubles compared to that in the canonical Benalcazar-Bernevig-Hughes model, signaling an enriched topological structure. Our study establishes a fundamental connection between two paradigmatic topological phases and paves the way for further exploring spin-orbit-coupling induced exotic higher-dimensional topological phases.
\end{abstract}
	
\maketitle
	
\section{Introduction}

Topological phases of matter have fundamentally reshaped our understanding of quantum materials, revealing exotic states characterized not by symmetry breaking, but by global topological invariants. Among them, Weyl semimetals and higher-order topological phases represent two distinct yet equally intriguing classes~\cite{1-1,1-2,1-3,1-4,1-5,1-6,1-7,1-8,1-9,1-10,1-11,1-12,1-13}. Topological Weyl semimetal phases host gapless bulk excitations, known as Weyl nodes, acting as momentum-space analogs of relativistic Weyl fermions, carrying quantized monopole charges of Berry curvature and giving rise to topologically protected Fermi arc surface states~\cite{1-1,1-2,1-3,1-4,1-5}. In contrast, higher-order topological phases exhibit gapless boundary modes localized on boundaries of codimension greater than one, extending the conventional notion of bulk-boundary correspondence~\cite{1-6,1-7,1-8,1-9,1-10,1-11,1-12,1-13}. For instance, a third-order topological phase in three-dimension could host localized modes at crystal hinges or corners despite gapped surface modes. Such topological phases emerges from a combination of symmetry protection and nontrivial higher-order band topology, often characterized by new classes of topological invariants. However, as these two phases exhibit fundamentally distinct topological features, the connection between them, particularly their coexistence or transitions within a single system, remains poorly understood.

Although topological phases were first explored in electronic systems~\cite{2-1,qi2011}, recent years have seen their remarkable extension to classical-wave or synthetic quantum platforms~\cite{3-2,3-3,3-4,3-5,3-6,3-7,3-8,3-9,3-10,3-11,3-12,3-13}. These systems offer unique advantages for studying topological physics, such as high controllability and flexible geometries, enabling the direct realization, manipulation, and visualization of topological phases. In particular, both topological Weyl semimetal phases and higher-order topological phases have been widely demonstrated in classical-wave systems, including photonics~\cite{4-1,4-2,4-3,4-4,4-5,4-6,4-7,4-8,4-9,4-10,4-11,4-12,4-13,4-14,4-15,4-16,4-17,4-18,4-19,4-20,4-21,4-22,4-23,4-24,4-25,4-26}, acoustics~\cite{5-1,5-2,5-3,5-4,5-5,5-6,5-7,5-8,5-9,5-10,5-12,5-13,5-14,5-15,5-16,5-17,5-18,5-19,5-20,5-21,5-22,5-23,5-24,5-25} and circuit metamaterials~\cite{6-1,6-2,6-3,6-4,6-5,6-6,6-7,6-8,6-9,6-10,6-12,6-13,6-14,6-15,6-16,6-17,6-18,6-19,6-20}. Despite these significant experimental advances, it remains a considerable challenge to implement spin-orbit couplings in synthetic systems~\cite{wang2012spin,huang2016,Zhu2006,shao2008}, especially in three dimensions. This mechanism is essential for realizing a wide range of exotic higher-dimensional topological phases of matter.

In this work, we experimentally realize three-dimensional spin-orbit couplings in circuit metamaterials, and observe both the coexistence of and the controllable switch between Weyl semimetal phases and third-order topological phases, despite the fundamentally distinct topological nature of these two phases. Through theoretical analysis of topological invariants and boundary states, we show that both phases coexist in a lattice model with dimerized three-dimensional spin-orbit couplings. By tuning the strength of the dimerization, this model features a direct topological transition between the Weyl semimetal phase and the third-order topological insulator phase. Experimentally, we implement the synthetic three-dimensional spin-orbit couplings in a three-dimensional circuit network, based on meticulously engineering the circuit nodes and interconnections. By measuring the admittance spectra and local density of states (LDOS), we successfully measure out the topological Fermi arcs and reveal the numbers of topological corner modes, which are hallmark signatures of topological Weyl semimetal phases and third-order topological insulator phases, respectively. Our experimental results reveal that the third-order topological phase induced by dimerized spin-orbit coupling hosts twice as many corner modes as the seminal Benalcazar-Bernevig-Hughes model, providing compelling evidence of a richer higher-order topology in our system.

The paper is organized as follows. Section~\ref{sec:II} presents the dimerized three-dimensional spin-orbit coupling lattice model hosting topological phase transition between topological Weyl semimetal and third-order topological insulator phases. Section~\ref{sec:III} introduces its experimental implementation in circuit metamaterials.
Section~\ref{sec:IV} explores the experimental observation of Weyl semimetal and third-order topological insulator phases. Section~\ref{sec:V} gives a summary and outlook for this paper.
	
\section{Dimerized three-dimensional spin-orbit coupling lattice model}
	\label{sec:II}
	\begin{figure*}[htbp]
		\includegraphics[width=0.85\textwidth,height=0.5\textwidth]{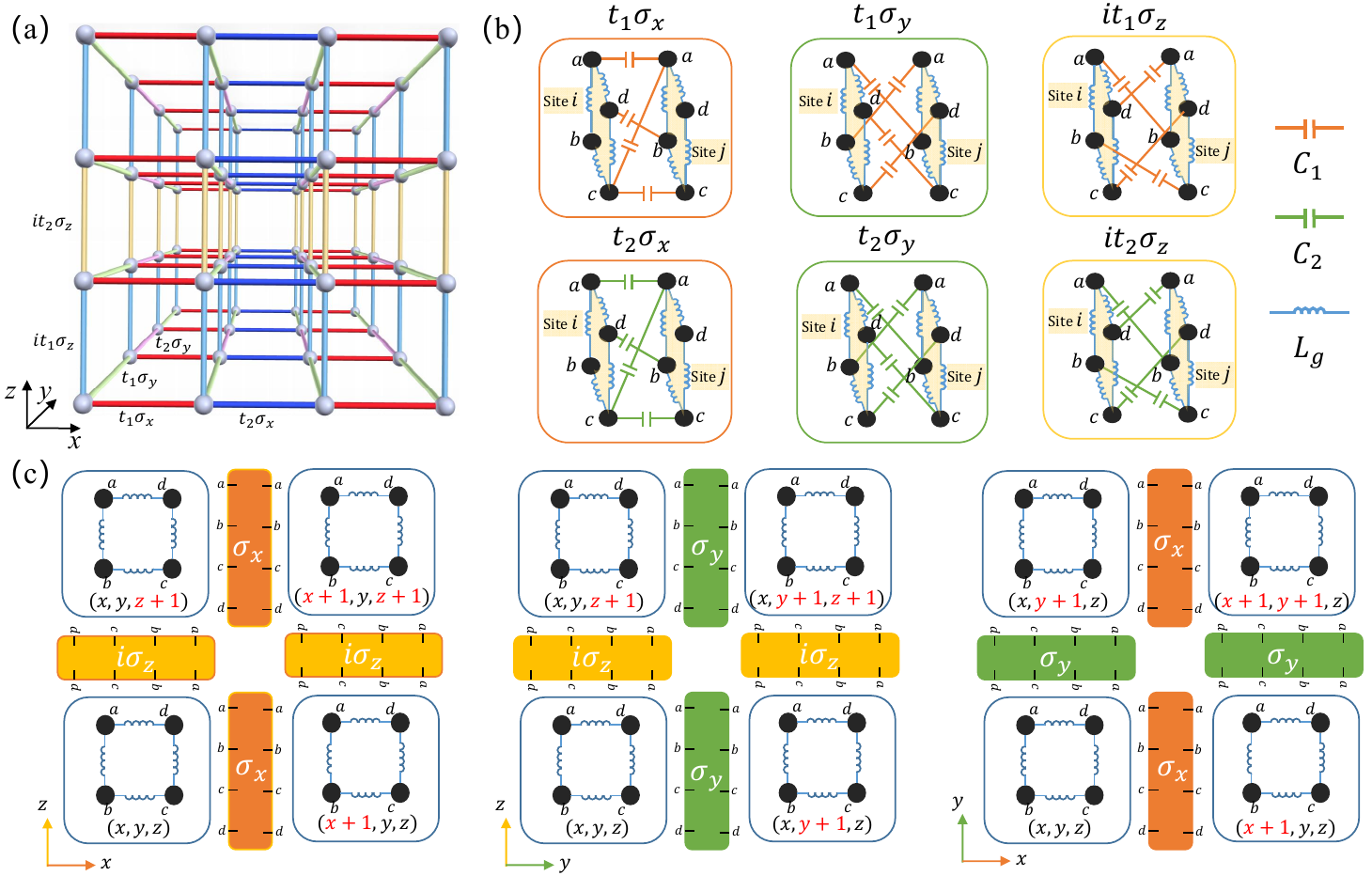}
		\caption{(a) Schematic of the lattice model with dimerized three-dimensional SOCs. Every lattice site denoted by gray ball contains a spin up and a spin down. $t_{1,2}$ are the intra- and inter-cell coupling strengths. (b) Circuit implementation of the SOCs within and between nearest-neighbour unit cells along three real-space directions. Each pseudospin is encoded and simulated by four circuit nodes connected through the inductances $L_g$, with SOCs realized by the connections through capacitors $C_{1,2}$. (c) Schematic diagram of circuit connections for implementing the three-dimensional lattice model in (a).}
		\label{Fig1}
	\end{figure*}
	As illustrated in Fig. \ref{Fig1}(a), we construct a three-dimensional lattice model with dimerized spin-orbit couplings along three spatial directions $x$, $y$, and $z$, where each lattice site contains two pseudospin degrees of freedom. The corresponding Bloch Hamiltonian in the momentum space takes the following form
	\begin{equation}
		\begin{aligned}
			H(\textbf{k})=&(t_1+t_2\cos(k_x)){\Gamma _1}+(t_1+t_2\cos(k_y)){\Gamma _2}\\&+(t_1-t_2\cos(k_z)){\Gamma _3}-t_2\sin(k_x){\Gamma _4}\\& - t_2\sin(k_y){\Gamma _5}+t_2\sin (k_z ){\Gamma _6},
		\end{aligned}
	\end{equation}
	where $t_{1,2}$ represent the inter- and intra-cell couplings, respectively. ${\Gamma _1}=\xi_0{\varsigma _x}{\tau _0}{\sigma _x}$, ${\Gamma _2}=\xi_0{\varsigma _x}{\tau _x}{\sigma _y}$, ${\Gamma _3}=\xi_y{\varsigma _0}{\tau _0}{\sigma _z}$, ${\Gamma_4}=\xi_0{\varsigma _y}{\tau _z}{\sigma _x}$, ${\Gamma _5}=\xi_0{\varsigma _x}{\tau _y}{\sigma _y}$, ${\Gamma _6}=\xi_x{\varsigma _0}{\tau _0}{\sigma _z}$, where $\xi_v$, $\tau_v$ and $\varsigma_v$ are the Pauli matrices defined by the sublattice degrees of freedom within the unit cell, and $\sigma_v$ is the Pauli matrix defined by the pseudospin degree of freedom. Specifically, for $t_1=t_2=t$, this three-dimensional lattice model is descried by the ideal topological Weyl semimetal Hamiltonian. The corresponding Bloch Hamiltonian reads as
	\begin{equation}
		H_w(k)=2t\left[\cos(k_x) \sigma_x+\cos(k_y)\sigma_y +\sin(k_z) \sigma_z\right].
		\label{H_Weyl}
	\end{equation}
	As shown, it has two energy bands, $E=\pm 2t\sqrt{\cos(k_x)^2+\cos(k_y)^2+\sin(k_z)^2}$. Within the first Brillouin zone, the two bands touch at eight points: $\textbf{k}_w=(k_x,k_y,k_z) = (\pm\pi/2, \pm\pi/2, 0)$ and $(\pm\pi/2, \pm\pi/2, \pi)$. In the following, we will show that these band-touching points are Weyl points characterized by topological Chern numbers.
	
	For this purpose, we first expand the Hamiltonian in Eq. \ref{H_Weyl} around the band-touching points. The resulting low-energy Hamiltonian can be expressed as $H_w = \sum_{ij} q_i v_{ij} \sigma_j$, where \(\textbf{q} = (q_x, q_y, q_z) = \textbf{k} - \textbf{k}_w\) and \(\textbf{k}\) is the momentum near the touching points. Here, \([v_{ij}]\) is a \(3 \times 3\) matrix with nonzero elements: $v_{xx} = \mp 2t$, $v_{yy} = \mp 2t$, $v_{zz} = \pm 2t$. It is evident that this Hamiltonian takes the form of a Weyl Hamiltonian, confirming that the band-touching points are indeed Weyl points. The topological nature of each Weyl point is characterized by its chirality, defined as $C=\text{sign}(v_{xx} v_{yy} v_{zz})$. In our system, the four Weyl points at $(\frac{\pi}{2},\frac{\pi}{2},0), (\frac{\pi}{2},-\frac{\pi}{2},\pi), (-\frac{\pi}{2},\frac{\pi}{2},\pi)$ and $(-\frac{\pi}{2},-\frac{\pi}{2},0)$ exhibit chirality $C=1$, while the four points at $(\frac{\pi}{2},\frac{\pi}{2},\pi),(\frac{\pi}{2},-\frac{\pi}{2},0),(-\frac{\pi}{2},\frac{\pi}{2},0)$, and $
	(-\frac{\pi}{2},-\frac{\pi}{2},\pi)$ have chirality $C=-1$. Weyl points can also be seen as monololes of Berry curvature in the momentum space, characterized by an integer-valued monopole charge. This charge corresponds to the Chern number computed over a closed two-dimensional surface in momentum space that encloses the Weyl point. Mathematically, the Chern number is given by $C=\frac{1}{2\pi}\oint_{S}\Omega(\textbf{q})\cdot dS$, where $\Omega(\textbf{q})=\nabla_\textbf{q}\times A(\textbf{q})$ is the Berry curvature and $\mathcal{A}(\textbf{q})=i\langle\psi(\textbf{q})|\nabla_{\mathbf{q}}|\psi(\textbf{q})\rangle$ is the Berry connection. Specifically, the Berry curvatures of the Weyl points with chiralities $1$ and $-1$ are calculated as $\textbf{q}/2|\textbf{q}|^3$ and $-\textbf{q}/2|\textbf{q}|^3$, respectively. Consequently, the associated Chern numbers are $C=1$ and $-1$, respectively.
	
	According to bulk-edge correspondence, the topological charge of the Weyl points is also reﬂected in the structure of the surface states. This is most easily understood by considering a simple system with only one pair of Weyl points and examining the states on an infinite cylindrical surface whose axis is aligned with the vector connecting the oppositely charged Weyl points. In this geometric configuration, the resulting two-dimensional band structure exhibits chiral modes propagating along the surface. The momentum structure of these surface states is model dependent, but the surface channel must terminate at the projections of the Weyl points onto the surface, yielding “Fermi arcs” that connect the Weyl points. We experimentally probe the surface physics by cleaving the three-dimensional lattice along the direction perpendicular to $\hat{v}=\hat{x}-\hat{y}$, while maintaining infinite extension along the  $\hat{z}$ and $\hat{u}=\hat{x}+\hat{y}$ directions. 
	
	In contrast, by tuning $t_1\ne t_2$, we can switch the Weyl semimetal phase into the third-order topological insulator phase. In such phase, the system possesses three sets of mirror symmetry operations along the $x$, $y$, and $z$ directions, expressed as:  ${M}_x=\xi_z{\varsigma _y}{\tau _z}{\sigma _y}$, ${M}_y=\xi_z{\varsigma _x}{\tau _y}{\sigma _x}$, ${M}_z=\xi_y{\varsigma _z}{\tau _0}{\sigma _z}$.  Crucially, these three mirror operations anticommute pairwise, satisfying $\{M_x, M_y\} = \{M_y, M_z\} = \{M_x, M_z\} = 0$. Due to the system remains invariant under mirror symmetry $M_j$, the Hamiltonian satisfies $[M_j, H(k)] = 0$, its eigenstates can be chosen as eigenstates of $M_j$: $M_j \ket{\mu_n(\mathbf{k})} = \lambda_n \ket{\mu_n(\mathbf{k})}$. Now consider another mirror operator $M_i$ that anticommutes with $M_j$ ($\{M_i, M_j\} = 0$), applying $M_i$  to $\ket{\mu_n(\mathbf{k})}$ yields: $ M_j \big( M_i \ket{\mu_n(\mathbf{k})} \big) = -M_i M_j \ket{\mu_n(\mathbf{k})} = -\lambda_n M_i \ket{\mu_n(\mathbf{k})} $. This shows that $M_i \ket{\mu_n(\mathbf{k})}$ is an eigenstate of $M_j$ with eigenvalue $-\lambda_n$, meaning the original $M_j$ eigenspace is mapped to an orthogonal subspace under $M_i$. This implies that the system cannot be simultaneously diagonalized under both $M_i$ and $M_j$, the two mirror operations do not share a common eigenbasis.
	Further considering a corner point $r$ located at the intersection of three mirror-symmetric planes ($x=0$, $y=0$, $z=0$), if the corner state $\ket{\phi_{\text{corner}}}$ simultaneously satisfies all three mirror symmetries: $M_x\ket{\phi} = \lambda_x\ket{\phi}$, $M_y\ket{\phi} = \lambda_y\ket{\phi}$, $M_z\ket{\phi} = \lambda_z\ket{\phi}$, then the anticommutation relations among them imply: $M_x M_y M_z \ket{\phi} = -M_y M_x M_z \ket{\phi} = -M_x M_y M_z \ket{\phi}$, leading to $\ket{\phi} = 0$. This demonstrates that no eigenstate can simultaneously satisfy all mirror symmetries unless the corner forms a special compatible subspace.  As a result, the algebraic structure of the mirror symmetries imposes a strong constraint on the existence of the corner states.
	
	Furthermore, the system we study possesses chiral symmetry defined by the operator $\Gamma = \xi_z  \varsigma_0  \tau_0 \sigma_z$, which constrains the Hamiltonian through the symmetry relation: $\Gamma H(\mathbf{k}) \Gamma^{-1} = -H(\mathbf{k})$. Under this symmetry protection, the system admits the definition of a multipole chiral number as a topological invariant. This topological invariant is constructed by defining octupole moment operators $Q_{xyz}^A$ and $Q_{xyz}^B$ on sublattices $A$ and $B$, respectively, and expressing their representations in their respective subspacess $U_A$ and $U_B$ as $\bar{O}_{xyz}^S = U_S^\dagger O_{xyz}^S U_S$, for $S = A, B$. The invariant is then expressed as: $N = \frac{1}{2\pi i} \Tr\log(\bar{Q}_{xyz}^A \bar{Q}_{xyz}^{B\dagger})$~\cite{chiral}, which determines the numbers of corner states as $4N$. In our system, when $t_1/t_2 > 1$, topological index $N$ identifies it as trivial ($N=0$); As $t_1/t_2$ decreases and crosses the critical point $t_1=t_2$, the system accompanied by a topological transition; For $t_1/t_2 < 1$, the topological index is $N=2$, showing that this system is in a nontrivial phase. Under open boundary conditions, two independent zero-energy modes localize at each of the eight crystalline corners, collectively forming a 16-fold degenerate corner states.
	
	\begin{figure*}[htbp]
		\includegraphics[width=1\textwidth,height=0.42\textwidth]{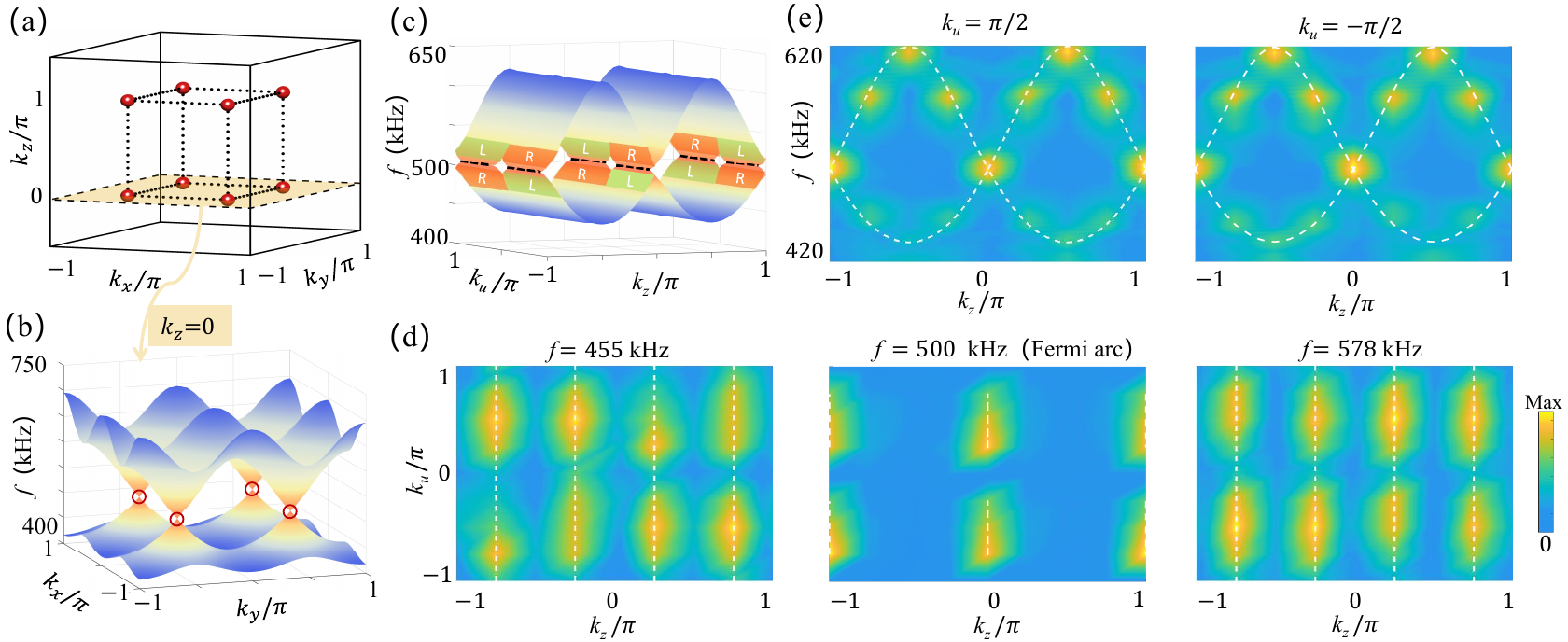}
		\caption{(a) Schematic of the first Brillouin zone and the eight represented Weyl points at $(k_x=\pm \pi/2,k_y=\pm \pi/2,k_z=0,\pi)$, with (b) exemplifying the four Weyl points in the low-energy spectrum of $k_z=0$ plane. (c) Simulated surface-mode eigenfrequency spectra for the circuit lattice with $\hat{u}=\hat{x}+\hat{y}$ and $\hat{z}$ under periodic boundary conditions and $\hat{v}=\hat{x}-\hat{y}$ under open boundary conditions. The two dispersion sheets $L$ (green) and $R$ (orange) represent the two surface modes, with the intersections at the circuit resonant frequency $f_0=500$ kHz being the synthetic topological Fermi arcs. (d) Measured (color map) and simulated (white dashed lines) surface-mode eigenfrequency spectrum for the circuit driven frequency $f=455$ kHz, $500$ kHz and $578$ kHz. (e) Measured and simulated surface-mode eigenfrequency spectrum for the momentum $k_u=\pm\pi/2$. The circuit parameters are $C_1=C_2=3.3$ nF and  $L_g=10$ $\mu$H. }
		\label{Fig2}
	\end{figure*}
	
	\section{Experimental implementation in circuit metamaterials}
	\label{sec:III}
	Circuit metamaterials~\cite{z-1,z-2,z-3,z-4}, composed of capacitors, inductors, and resistors, provide highly tunable parameters and direct spectral access via impedance measurements, enabling the simulation of both Hermitian and non-Hermitian topological states~\cite{helbig2020,N-4}. Building on this platform, we designed a circuit metamaterial to emulate a dimerized three-dimensional spin–orbit-coupled lattice model, where the arrangement of discrete-element modules encodes pseudospin and spin–orbit couplings~\cite{wu2022}, thereby capturing the key physical mechanisms of the model, as illustrated in Fig. \ref{Fig1}(b). Each pseudospin circuit module consisting of four nodes, labeled $a$, $b$, $c$, and $d$, arranged in a cyclic quadrilateral and interconnected by identical inductors $L_g$. Specifically, the voltage signals $V_{i,a}$, $V_{i,b}$, $V_{i,c}$, and $V_{i,d}$ at each four-node module can be linearly combined to construct the pseudospin: $ V_{i,\uparrow} = V_{i,a} +e^{i\pi/2}V_{i,b}  +e^{i\pi}V_{i,c} +e^{i3\pi/2}V_{i,d} $ and $ V_{i,\downarrow} = V_{i,a} +e^{i3\pi/2}V_{i,b} +e^{i\pi}V_{i,c} +e^{i\pi/2}V_{i,d} $. Three types of SOCs $\sigma_x$, $\sigma_y$ and $i\sigma_z$ are realized through engineered node-to-node connections: $ \sigma_x $ coupling connects nodes [$i_a,\ i_b,\ i_c,\ i_d$] to $\left[ j_a,\ j_d,\ j_c,\ j_b \right]$, $\sigma_y$ coupling maps to $\left[ j_b,\ j_a,\ j_d,\ j_c \right]$, and $i\sigma_z$ coupling links to $\left[j_b,\ j_c,\ j_d,\ j_a\right]$ (see Supplemental Material for more details). Capacitors $C_1$ and $C_2$ correspond to intra-cell and inter-cell hopping strengths $t_1$ and $t_2$, respectively.  Building upon this foundation, the coupling structure of the basic unit cells used to construct the three-dimensional circuit lattice in each plane is shown in Fig. \ref{Fig1}(c), which specifically illustrates the SOCs of the pseudospin modules along different spatial directions. 
	
	We derive the eigen-equation of the designed electric circuit metamaterial based on Kirchhoff’s law to verify its validity. According to Kirchhoff’s law, the external current flowing into node $\alpha$ at frequency $\omega$ is given by $I_\alpha(\omega)=\sum_{\beta}{J_{\alpha \beta}(\omega,k)V_\beta(\omega)}$, where $I_\alpha(\omega)$ denotes the input current at node $\alpha$, $V_\beta(\omega)$ is the voltage at node $\beta$, and $J_{\alpha\beta}(\omega, k)$ represents the element of the circuit Laplacian matrix. Here, $k$ denotes the crystal momentum in the circuit's synthetic momentum space, and $\omega=2\pi f$ is the AC driving frequency. In the pseudospin basis, the Kirchhoff equations take the compact matrix form: $\tilde J(\omega,k)\tilde V(\omega) =\tilde I(\omega)$, where $\tilde V(\omega) =[V_{\uparrow}(\omega),V_{\downarrow}(\omega)]^T$ and $\tilde I(\omega) =[I_{\uparrow}(\omega),I_{\downarrow}(\omega)]^T$ are the voltage and current vectors in the pseudospin space, respectively. In the absence of external current input, e.g. $\tilde I(\omega)=0$, impedance equation of the SOC circuit metamaterial is
	\begin{equation}
		\begin{aligned}
			E(\omega)\tilde V(\omega)=& i\omega[( C_1+ C_2\cos(k_x)){\Gamma _1}+(C_1+ C_2\cos(k_y)){\Gamma _2}\\&+( C_1- C_2\cos(k_z)){\Gamma _3}-t_2\sin(k_x){\Gamma _4}\\& - C_2\sin(k_y){\Gamma _5}+ C_2\sin (k_z ){\Gamma _6} ],
		\end{aligned}
	\end{equation}
	where $E(\omega)=3i\omega(C_1+C_2)+2/(i\omega L_g)$. The detail derivation can be found in the Supplemental Material. Hence, it is straightforward to infer that our designed electric circuit metamaterials can implement the dimerized 3D SOCs lattice model.
	
	\section{Experimental observation of Weyl semimetal and third-order topological insulator phases}
	\label{sec:IV}
	\begin{figure*}[htbp]
		\includegraphics[width=1\textwidth,height=0.5\textwidth]{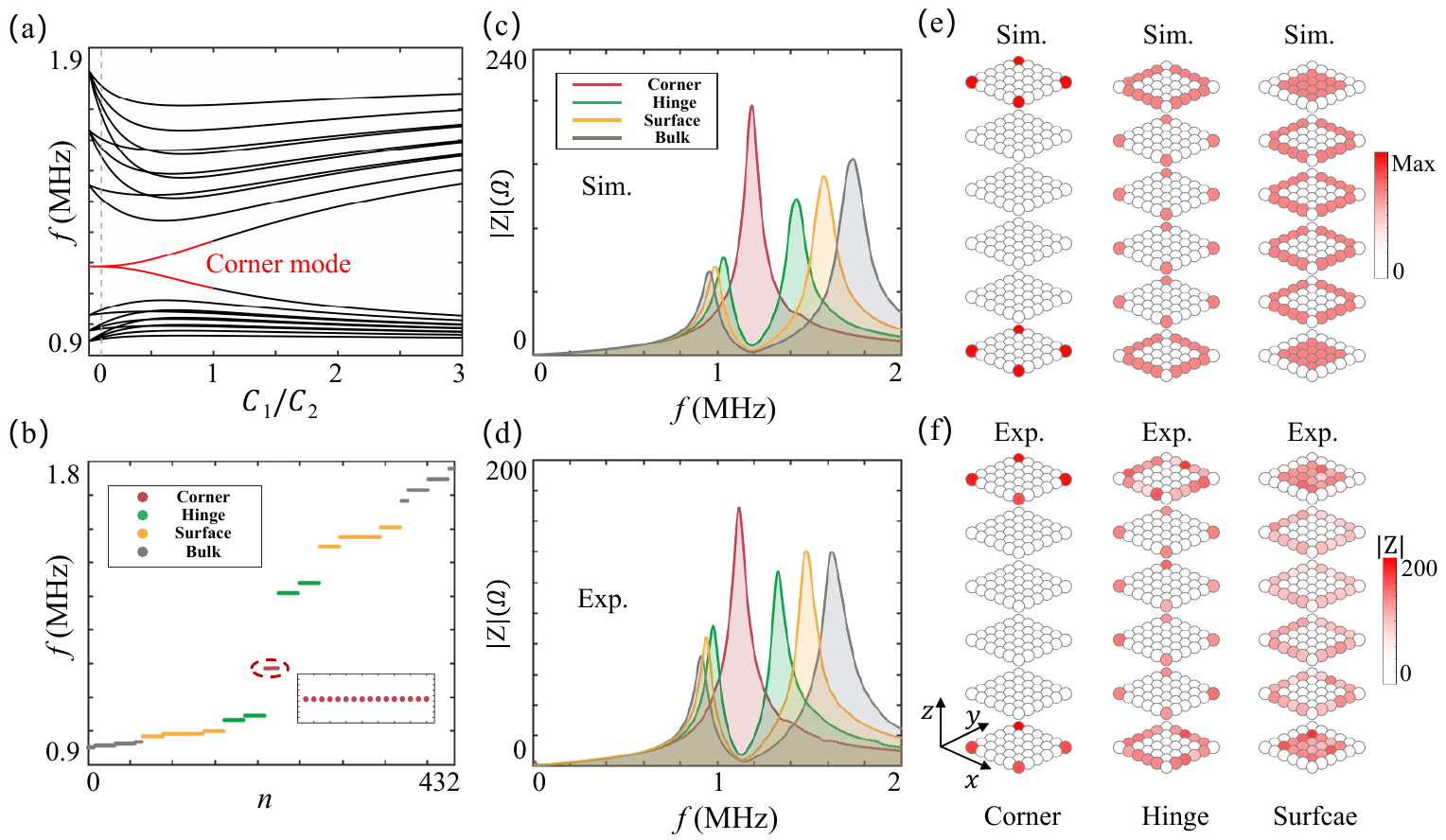}
		\caption{(a) Simulated eigenfrequency spectrum of the circuit lattice as a function of $C_1/C_2$ under open boundary conditions along the three directions. The red lines represent the topological corner modes and the non-degeneracy is caused by the finite-size effect. (b) Simulated eigenfrequency spectrum for $C_1/C_2=0.1$, where the red, green, yellow, and gray curves represent the corner, hinge, surface, and bulk modes, respectively. Notably, there are sixteen degenerate topological corner modes in the spectrum gap. (c) Simulated and (d) measured impedances between the corner, hinge, surface, bulk nodes and the grounding nodes as a function of the driving frequency. (e) Simulated density distribution of the topological corner, hinge and surface modes. (f) Measured impedance distributions for $f=1.11$ MHz, $1.33$ MHz and $1.48$ MHz, corresponding to the topological corner, hinge and surface modes, respectively. The circuit parameters are $C_1=330$ pF, $C_2=3.3$ nF and $L_g=3.3$ $\mu$H.}
		\label{Fig3}
	\end{figure*}
	
	We first consider the zero-dimerized case by setting the circuit parameters to $C_1 = C_2 = 3.3$ nF and $L_g = 10$ $\mu$H, in order to observe the topological Weyl semimetal behavior in the constructed 3D SOC circuit system. At the resonant frequency $f_0 = 1/(2\pi\sqrt{3C_1L_g}) \approx 500$ kHz, the system hosts eight Weyl points under periodic boundary conditions, as shown in Fig. \ref{Fig2}(a). Among them, four Weyl points lie within the $k_z = 0$ plane [Fig. \ref{Fig2}(b)], emerge from linear band crossings in momentum space. To experimentally observe the topological Fermi arc surface modes, we fabricated a 3D SOC circuit lattice consisting of $8 \times 8 \times 8$ unit cells using a stacked printed circuit board (PCB) architecture. The lattice is constructed with translational vectors along the $\hat{u}$, $\hat{v}$, and $\hat{z}$ directions. In Weyl semimetal, the existence of topological Fermi arcs depends on whether the projected chiral charges of Weyl points cancel on the observed surface. Since projected Weyl points with opposite chirality may annihilate on certain surfaces, we selected the $\hat{u}$–$\hat{z}$ plane—implementing periodic boundary conditions along the $\hat{u}$ and $\hat{z}$ directions via PCB traces and external wiring, while leaving the $\hat{v}$ direction open—to preserve their separation and enable the observation of topological Fermi arc. The simulated surface-mode eigenfrequency spectra on this plane is shown in Fig. \ref{Fig2}(c), revealing six distinct topological Fermi arcs that connect the projected Weyl nodes of opposite chirality near $f_0 = 500$ $\mathrm{kHz}$.
	
	The circuit metamaterial was experimentally characterized using a vector network analyzer (VNA) Keysight E5061B, which enabled simultaneous excitation and measurement of frequency responses. A frequency-swept voltage signal ranging from $420$ kHz to $620$ kHz was applied to each surface node in the $u$–$z$ plane, and the voltage responses were recorded at all 64 detection points across the circuit surface. The momentum-space spectral function was reconstructed via a two-dimensional discrete Fourier transform: $\tilde{P}_{k_u,k_z}(\omega) = \sum_{u,z=0}^7 P_{u,z}(\omega) e^{i[uk_u + zk_z]}$, where $P_{u,z}(\omega)$ denotes the two-point impedance between sites indexed by $u$ and $z$, and $k_u$, $k_z = 2\pi m / 8$ with $m = 0, 1, \ldots, 7$ represent the discrete lattice momenta along the $\hat{u}$ and $\hat{z}$ directions, respectively. Fig. \ref{Fig2}(d) shows the surface-mode impedance spectra at three representative driving frequencies-$455$ kHz, $500$ kHz, and $578$ kHz-near the resonant frequency $f_0$, with the experimental and simulated results represented by the color maps and white dashed lines, respectively. At $455$ kHz and $578$ kHz, the spectra exhibit four dispersive lines across $k_z$-$k_u$ plane, consistent with the presence of gapped surface modes. In contrast, at $500$ kHz, six topological Fermi arcs connecting Weyl points are clearly resolved at $k_z = 0$ and $\pi$, demonstrating excellent agreement between experiment and theory. To further examine the surface-mode dispersion, we extracted the impedance spectra along the $k_z$ direction at fixed momenta $k_u = \pm \pi/2$. The resulting impedance curves, shown in Fig. \ref{Fig2}(e), clearly reveal linearly crossing surface modes that form topological Fermi arcs in momentum space, in excellent agreement with theoretical predictions and providing further confirmation of the surface topological states.
	
	\begin{figure*}[htbp]
		\includegraphics[width=1\textwidth,height=0.45\textwidth]{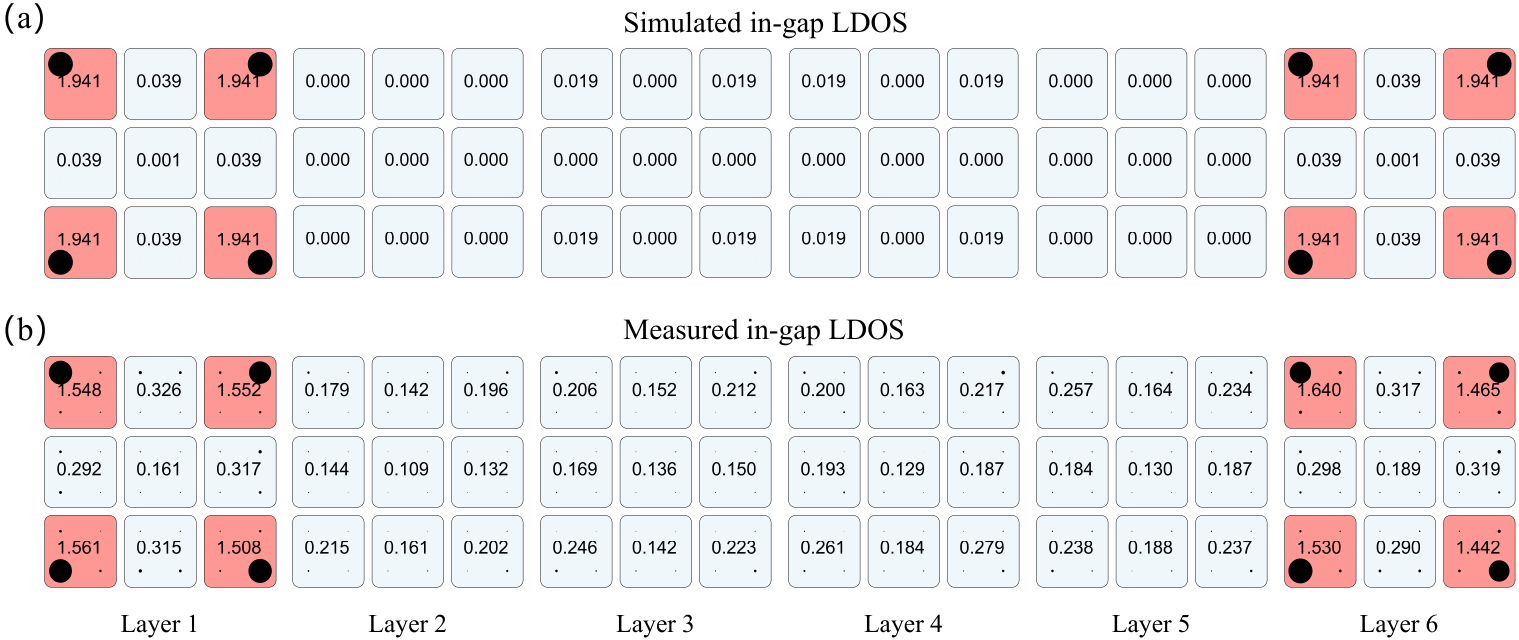}
		\caption{Spatial maps of simulated (a) and measured (b) in-gap LDOS. Each circuit site is represented as a circle with radius proportional to the LDOS. Each number gives the LDOS in the corresponding unit cell. As shown, the predominated LDOS in the corner unit cells of the top and bottom layers clearly manifest the emergence of topological corner modes there.}
		\label{Fig4}
	\end{figure*}
	
	We next consider the dimerized case to simulate a three-dimensional third-order topological insulator phase, using a system composed of $3\times3\times3$ circuit unit cells. To verify the topological nature of this configuration, we first carried out a theoretical analysis of the system's eigenfrequency spectra evolution as a function of the capacitor ratio $C_1/C_2$. Setting $C_1 = 330$ pF and resonant frequency $f_0 = 1.19$ MHz, the simulated eigenfrequency spectra as $C_1/C_2$ varies continuously are shown in Fig. \ref{Fig3}(a). A distinct bandgap opens between the main bands for $0 < C_1/C_2 < 1$, within which isolated corner modes (red lines) appear—signaling that the system enters a third-order topological phase. For detailed characterization, we selected a representative parameter $C_1/C_2 = 0.1$ and plotted the full eigenfrequency distribution in Fig. \ref{Fig3}(b). Under this setting, the eigenfrequency spectrum exhibits clear band separation, with eigenmodes color-coded based on their spatial localization: bulk modes (gray), surface modes (yellow), hinge modes (green), and corner modes (red). The corner modes appear the center of the main bandgap and are eigenfrequency well isolated from the other modes, providing a clear signature of higher-order topology. The 16 observed corner modes are distributed over the eight geometric corners of the 3D lattice, with two localized corner modes per corner, corresponding precisely to the multipole chiral number $N = 2$ and confirming the realization of the third-order topological phase. The distinct spectral location of these modes also establishes a reliable frequency-domain reference for subsequent experimental identification. 
	
	In the experiment, the hallmark features of the third-order topological insulator phase—namely, the corner modes impedance peak at resonant frequency and their spatial density distributions—are probed by measuring the impedance $Z_a$ between node $a$ and ground. This impedance is directly related to the eigenvalues and eigenvectors of the circuit Laplacian, $j_n$ and $\psi_n$, through the relation $Z_{a}(\omega)=\frac{\tilde V_a(\omega)}{\tilde I_{a}(\omega)}=\sum\limits_{n}{\frac{|\psi_{n,a}|^2}{j_n(\omega)}}$. This expression implies that when the driving frequency $\omega$ matches a circuit eigenfrequency, the corresponding mode's impedance response is strongly enhanced, resulting in observable resonance peaks in the frequency spectrum.
	Fig. \ref{Fig3}(c) verifies this feature based on simulating the measurements using the LTspice simulation software. As expected, the largest impedance peaks at corner sites appear near the resonant frequency $f_0 = 1.19$ MHz, while hinge, surface, and bulk sites also exhibit peaks at their respective eigenfrequencies. This phenomenon confirms the effective frequency-domain separation of boundary modes of different orders. Experimental measurements of the corner, hinge, surface, and bulk impedances are shown in Fig. \ref{Fig3}(d). Despite minor frequency shifts caused by the 5\% tolerance in circuit components, all measured responses show excellent agreement with simulation results. To verify the real-space localization characteristics of the boundary modes, we simulated the spatial density distributions of the topological corner, hinge, and surface modes, as shown in Fig. \ref{Fig3}(e). The results reveal that the corner modes are confined to the lattice vertices, hinge modes are distributed along the edges, and surface modes are concentrated on the faces, collectively providing direct evidence for the third-order topological nature of the system. In the experiment, we measured the impedance distributions at three representative frequencies, $1.11$ MHz for corner modes, $1.33$ MHz for hinge modes, and $1.48$ MHz for surface modes, as shown in Fig. \ref{Fig3}(f). The measured results clearly reveal the localization of boundary modes at the lattice corners, edges, and surfaces, in excellent agreement with theoretical predictions.
	
	To characterize the third-order topological insulator phase in the 3D SOC system, it is desirable to determine the number of corner modes. Although the corner modes are degenerate in the eigenfrequency spectrum and cannot be directly resolved, we show that their number can still be inferred from LDOS measurements~\cite{peterson2020,peterson2021,liu2021,xie2023,LDOS}. In circuit systems, the LDOS (mode density) at each node—for example, node $a$—is obtained from the real part of the impedance via the relation $\rho(f, a) = 2f\ \mathrm{Re}[Z_a]$ (see Supplemental Material for more details). To extract the total number of corner modes, we integrate the LDOS of each unit cell over an in-gap frequency range that encompasses corner modes contributions. The experimentally measured LDOS (Fig. \ref{Fig4}(b)) agrees well with the theoretical simulation (Fig. \ref{Fig4}(a)). The number of corner modes can be estimated by summing the in-gap LDOS across all unit cells. Although parasitic resistances and component tolerances lead to deviations between experiment and theory—making it difficult to precisely quantify the total in-gap LDOS and extract the exact number of corner modes—the spatial distribution of the LDOS provides clear evidence. Since the in-gap LDOS arises solely from corner modes, all non-corner unit cells exhibit negligible LDOS values, while only the corner sites contribute significantly. Specifically, the LDOS values at all corner unit cells remain close to $\rho \approx 2$, confirming the presence of twofold-degenerate corner modes at each of the eight corners. 
	
	\section{Summary and Outlook}
	\label{sec:V}
	In summary, we have experimentally realized three-dimensional spin-orbit couplings in circuit metamaterials, which is the key ingredient of implementing a variety of three-dimensional topological phases. In our work, we find such spin-orbit coupling could enable the coexistence of and controllable switching between Weyl semimetal and third-order topological phases. By measuring admittance spectra and local density of states, we observe key signatures of both topological phases, including the topological Fermi arcs and corner modes. Notably, the third-order phase exhibits a doubled corner-mode degeneracy compared to that in the seminal Benalcazar-Bernevig-Hughes model, providing compelling evidence of a richer higher-order topology induced by the spin-orbit couplings in our system. Looking ahead, our experiment is expected to inspire both future theoretical and experimental studies on the three-dimensional topological phases of matter emerged from the interplay between three-dimensional spin-orbit couplings and crystal symmetry. For example, it would be quite interesting to study how such interplay further enriches three-dimensional topological Dirac semimetal phases~\cite{7-1,7-2,7-3,7-4} and higher-order topological semimetal phases~\cite{8-1,8-2,8-3,8-4,8-5,8-6,8-7,8-8}.
	
	\section*{Acknowledgment}
	
	This work is supported by the by the National Key Research and Development Program of China (Grant No. 2022YFA1404201), National Natural Science Foundation of China (NSFC) (Grant No. 12474361, No. 12034012, No. 12074234), Changjiang Scholars and Innovative Research Team in University of Ministry of Education of China (PCSIRT)($IRT\_17R70$), Fund for Shanxi 1331 Project Key Subjects Construction, 111 Project (D18001) and Fundamental Research Program of Shanxi Province (Grant No. 202303021223005). Y.L. acknowledges support from the China Association for Science and Technology (CAST), China.
	
	\bibliographystyle{naturemag}
	\bibliography{ref.bib}

	\vspace{2em}
	
	\appendix
	
	\section*{ APPENDIX A: PSEUDO-SPIN SPACE IN CIRCUIT METAMATERIALS}
	\label{app:A}
	\setcounter{equation}{0}
	\renewcommand{\theequation}{A\arabic{equation}}
	
	\makeatletter
	\renewcommand{\theHequation}{A\arabic{equation}}
	\makeatother
	
	\begin{figure*}[htbp]	\includegraphics[width=1\textwidth,height=0.17\textwidth]{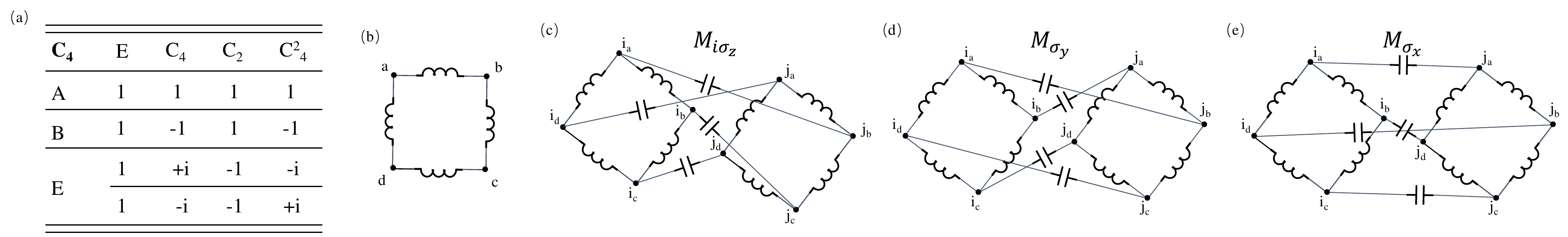}
		\caption{(a) Character table for $C_4$ point group. For systems with time-reversal symmetry, $E$ is twofold degenerate. (b) The pseudo-spin module $M_0$: four inductors with the same parameters are connected head to tail to form a square. The nodes are denoted as $ a, b, c$, and $d$. (c-e) SOC terms $i\sigma_z,\sigma_y,\sigma_x$ between two pseudo-spin modules cell-$i$ and cell-$j$, realized via module $M_{i\sigma_z}$, $M_{\sigma_y}$, and $ M_{\sigma_x}$, respectively.}
		\label{S1}
	\end{figure*}
	
	For the circuit shown in Fig. \ref{S1}(b), four identical inductors are connected head-to-tail to form a square, exhibiting $C_4$ rotational symmetry. As illustrated in Fig.  \ref{S1}(a), the $C_4$ symmetry admits a two-dimensional irreducible representation, whose basis functions are complex conjugates of each other. Accordingly, the doubly degenerate eigenstates of the circuit can be selected as the basis states of a pseudo-spin space. Taking this inductor circuit as an example, we verify this construction by analyzing its eigenmodes. The Kirchhoff current equations for the configuration in Fig. \ref{S1}(b) are given by:
	\begin{equation}
		\left( {\begin{array}{*{20}{c}}{{I_a}}\\{{I_b}}\\{{I_c}}\\{{I_d}}\end{array}} \right) 
		= \frac{1}{i\omega L}\left( {\begin{array}{*{20}{c}}{-2}&1&0&1\\1&{-2}&1&0\\0&1&{-2}&1\\1&0&1&{-2}\end{array}} \right)
		\left( {\begin{array}{*{20}{c}}{{V_a}}\\{{V_b}}\\{{V_c}}\\{{V_d}}\end{array}} \right)
		\label{S1_eq}
	\end{equation}
	where $V_{a,b,c,d}$ denote the voltages at the four nodes and $ I_{a,b,c,d}$ represent the currents flowing into each node. Here, $L$ is the inductance of each inductor, and $\omega$ is the frequency of the AC signal. We denote the $4 \times 4$ matrix appearing on the right-hand side of Eq. \ref{S1_eq} as $M_0$. By applying a unitary transformation $F$, we diagonalize $M_0$ as:
	\begin{equation}
	\begin{split}
		F = \frac{1}{2}\left( {\begin{array}{*{20}{c}}1&1&1&1\\1&{-1}&1&{-1}\\1&i&{-1}&{-i}\\1&{-i}&{-1}&i\end{array}} \right),\
		F{M_0}{F^{ - 1}} = -\left( {\begin{array}{*{20}{c}}0&0&0&0\\0&4&0&0\\0&0&2&0\\0&0&0&2\end{array}} \right)
			\label{S2_eq}
	\end{split}
	\end{equation}
	The twofold degenerate eigenstates in Eq. \ref{S2_eq}, namely, $ \phi_1=\frac{1}{2}(1,i,-1,-i)^T$ and $ \phi_2=\frac{1}{2}(1,-i,-1,i)^T$, serve as the basis functions of the pseudo-spin space, on which we will focus in the following analysis.
	
	To implement spin-orbit coupling terms $ \sigma_x$, $\sigma_y$ and $i\sigma_z$, we design capacitor networks that bridge different pseudo-spin modules. As an illustrative example, we consider the realization of the $i\sigma_z$ term, as depicted in Fig. \ref{S1}(c). According to Kirchhoff’s law, the current-voltage relation is given by $I(\omega)=J(\omega)V(\omega)$:
	\begin{widetext}
	\begin{equation} 
		\left( {\begin{array}{*{20}{c}}
				{{I_{i,a}}}\\{{I_{i,b}}}\\{{I_{i,c}}}\\{{I_{i,d}}}\\{{I_{j,a}}}\\{{I_{j,b}}}\\{{I_{j,c}}}\\{{I_{j,d}}}
		\end{array}} \right) = \left( {\begin{array}{*{20}{c}}
				{-2l-c}&l&0&l&0&c&0&0\\l&{-2l-c}&l&0&0&0&c&0\\0&l&{-2l-c}&l&0&0&0&c\\l&0&l&{-2l-c}&c&0&0&0\\0&0&0&c&{-2l-c}&l&0&l\\c&0&0&0&l&{-2l-c}&l&0\\0&c&0&0&0&l&{2l-c}&l\\0&0&c&0&l&0&l&{-2l-c}
		\end{array}} \right)\left( {\begin{array}{*{20}{c}}
				{{V_{i,a}}}\\{{V_{i,b}}}\\{{V_{i,c}}}\\{{V_{i,d}}}\\{{V_{j,a}}}\\{{V_{j,b}}}\\{{V_{j,c}}}\\{{V_{j,d}}}\end{array}} \right)
			\label{S3_eq}
	\end{equation}
	\end{widetext}
where $ l=\frac{1}{i\omega L} $ and $ c=i\omega C $ are the admittances of the inductors and capacitors, respectively, in the steady-state analysis of an AC signal. For convenience, we simplify the Laplacian matrix by defining the matrix:
\begin{equation} 
			{M_{i{\sigma _z}}} = \left( {\begin{array}{*{20}{c}}
				0&1&0&0\\
				0&0&1&0\\
				0&0&0&1\\
				1&0&0&0
		\end{array}} \right),\,
		{I_{4 \times 4}} = \left( {\begin{array}{*{20}{c}}
				1&0&0&0\\
				0&1&0&0\\
				0&0&1&0\\
				0&0&0&1
		\end{array}} \right)
\end{equation}
The Eq. \ref{S3_eq} becomes:
	\begin{equation}
		\left( {\begin{array}{*{20}{c}}{{{I}_i}}\\{{{I}_j}}
		\end{array}} \right) = \left( {\begin{array}{*{20}{c}}
				{2l{M_0} - c{I_{4 \times 4}}}&{c{M_{i{\sigma _z}}}}\\
				{cM_{i{\sigma _z}}^T}&{2l{M_0} - c{I_{4 \times 4}}}
		\end{array}} \right)\left( {\begin{array}{*{20}{c}}
				{{{V}_i}}\\	{{{V}_j}}\end{array}} \right)
	\end{equation}
	where $I_{i(j)}(\omega)=[I_{i(j),a}(\omega),I_{i(j),b}(\omega),I_{i(j),c}(\omega),I_{i(j),d}(\omega)]^T$ and $V_{i(j)}(\omega)=[V_{i(j),a}(\omega),V_{i(j),b}(\omega),V_{i(j),c}(\omega),V_{i(j),d}(\omega)]^T$ denote the current and voltage vectors for the $i$-th (or $j$-th) pseudo-spin module. Applying the unitary transformation matrix $F$ defined in Eq. \ref{S2_eq}, we transform into the pseudo-spin basis and obtain:
	\begin{equation}
		F{M^T_{i{\sigma _z}}}{F^{-1}} = \left( {\begin{array}{*{20}{c}}
				{1}&0&0&0\\0&-1&0&0\\0&0&i&0\\0&0&0&{-i}
		\end{array}} \right) = \left( {\begin{array}{*{20}{c}}
				{1}&0\\0&-1\end{array}} \right)\oplus i{\sigma _z}
	\end{equation}
	For the two degenerate $C_4$ symmetry-adapted basis functions defined in Eq. \ref{S2_eq}, the tunneling from module-$j$ to module-$i$ is characterized by the coupling matrix $i\sigma_z$. Using the same design principle, one can construct capacitor networks that realize $\sigma_x$ (Fig. \ref{S1}(d)) and $\sigma_y$ (Fig. \ref{S1}(e)) type coupling matrices between adjacent pseudo-spin modules:
		\begin{widetext}
		\begin{equation} 
		\begin{array}{l}
			{M_{{\sigma_x}}} = \left({\begin{array}{*{20}{c}}
					1&0&0&0\\0&0&0&1\\0&0&1&0\\0&1&0&0
			\end{array}}\right),F{M^T_{{\sigma_x}}}{F^{-1}}=\left({\begin{array}{*{20}{c}}
					1&0&0&0\\0&1&0&0\\0&0&0&1\\0&0&1&0
			\end{array}}\right)=\left({\begin{array}{*{20}{c}}1&0\\0&1\end{array}}\right)\oplus{\sigma_x}
			
			\\
			
			{M_{{\sigma_y}}}=\left({\begin{array}{*{20}{c}}
					0&1&0&0\\1&0&0&0\\0&0&0&1\\0&0&1&0
			\end{array}}\right),F{M^T_{{\sigma_y}}}{F^{-1}}=\left({\begin{array}{*{20}{c}}
					1&0&0&0\\0&{-1}&0&0\\0&0&0&i\\0&0&{-i}&0\end{array}}\right)=
			\left( {\begin{array}{*{20}{c}}1&0\\0&-1\end{array}} \right)\oplus{\sigma_y}
		\end{array}
		\end{equation}
	\end{widetext}
	Based on the above theoretical analysis, we demonstrate that input voltage signals with specific phase configurations—namely $\phi_1$ and $\phi_2$—can effectively excite the pseudo-spin degrees of freedom. Furthermore, the engineered inter-module coupling gives rise to well-defined spin-orbit interaction characteristics in the pseudo-spin space.
	
	\section*{APPENDIX B: DETAIL OF 3D SOC CIRCUIT METAMATERISLS}
	\begin{figure*}
		\includegraphics[width=0.3\textwidth,height=0.4\textwidth]{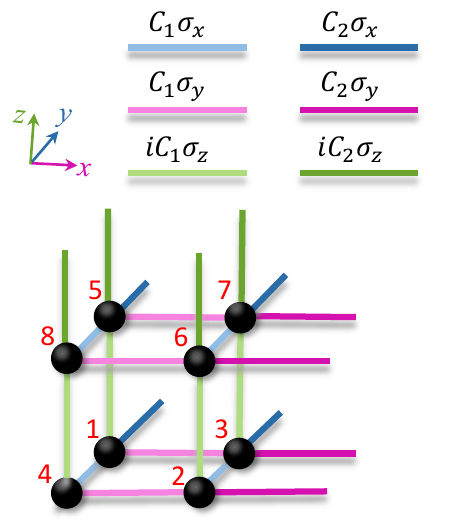}
		\caption{Model diagram for 3D SOC circuit metamaterials}
		\label{S2}
	\end{figure*}
	\setcounter{equation}{0}
	\renewcommand{\theequation}{B\arabic{equation}}
	
	\makeatletter
	\renewcommand{\theHequation}{B\arabic{equation}}
	\makeatother
	
	In this section, we provide the detailed construction of the three-dimensional spin–orbit coupling (3D SOC) circuit  metamaterials and derive the corresponding model Hamiltonian. For a 3D SOC circuit metamaterial under periodic boundary conditions, the application of Bloch’s theorem allows us to express Kirchhoff’s equations for sites 1 through 8 as:
	\begin{widetext}
		\begin{equation} 
			\begin{aligned}
			{I_1}& = ({l_g}{M_0} - 3({c_1} + {c_2}){I_{4 \times 4}}){V_1} + {c_1}({M_{{\sigma _x}}}{V_3} + {M_{{\sigma _y}}}{V_4} + {M_{i{\sigma _z}}}{V_5}) + {c_2}({e^{ikx}}{M}_{{\sigma _x}}^T{V_3} + {e^{iky}}{M}_{{\sigma _y}}^T{V_4} + {e^{-ikz}}{M}_{i{\sigma _z}}^T{V_5})\\			
			{I_2} &= ({l_g}{M_0} - 3({c_1} + {c_2}){I_{4 \times 4}}){V_2} + {c_1}({M_{{\sigma _x}}}{V_4} + {M_{{\sigma _y}}}{V_3} + {M_{i{\sigma _z}}}{V_6}) + {c_2}({e^{ - ikx}}{M}_{{\sigma _x}}^T{V_4} + {e^{ - iky}}{M}_{{\sigma _y}}^T{V_3} + {e^{-ikz}}{M}_{i{\sigma _z}}^T{V_6})\\
			{I_3} &= ({l_g}{M_0} - 3({c_1} + {c_2}){I_{4 \times 4}}){V_3} + {c_1}({M}_{{\sigma _x}}^T{V_1} + {M}_{{\sigma _y}}^T{V_2} + {M_{i{\sigma _z}}}{V_7}) + {c_2}({e^{ - ikx}}{M_{{\sigma _x}}}{V_1} + {e^{iky}}{M_{{\sigma _y}}}{V_2} + {e^{-ikz}}{M}_{i{\sigma _z}}^T{V_7})\\
			{I_4} &= ({l_g}{M_0} - 3({c_1} + {c_2}){I_{4 \times 4}}){V_4} + {c_1}({M}_{{\sigma _x}}^T{V_2} + {M}_{{\sigma _y}}^T{V_1} + {M_{i{\sigma _z}}}{V_8}) + {c_2}({e^{ikx}}{M_{{\sigma _x}}}{V_2} + {e^{-iky}}{M_{{\sigma _y}}}{V_1} + {e^{-ikz}}{M}_{i{\sigma _z}}^T{V_8})\\
			{I_5} &= ({l_g}{M_0} - 3({c_1} + {c_2}){I_{4 \times 4}}){V_5} + {c_1}({M_{{\sigma _x}}}{V_7} + {M_{{\sigma _y}}}{V_8} + {M}_{i{\sigma _z}}^T{V_1}) + {c_2}({e^{ikx}}{M}_{{\sigma _x}}^T{V_7} + {e^{iky}}{M}_{{\sigma _y}}^T{V_8} + {e^{  ikz}}{M_{i{\sigma _z}}}{V_1})\\
			{I_6} &= ({l_g}{M_0} - 3({c_1} + {c_2}){I_{4 \times 4}}){V_6} + {c_1}({M_{{\sigma _x}}}{V_8} + {M_{{\sigma _y}}}{V_7} + {M}_{i{\sigma _z}}^T{V_2}) + {c_2}({e^{ - ikx}}{M}_{{\sigma _x}}^T{V_8} + {e^{ - iky}}{M}_{{\sigma _y}}^T{V_7} + {e^{  ikz}}{M_{i{\sigma _z}}}{V_2})\\
			{I_7} &= ({l_g}{M_0} - 3({c_1} + {c_2}){I_{4 \times 4}}){V_7} + {c_1}({M}_{{\sigma _x}}^T{V_5} + {M}_{{\sigma _y}}^T{V_6} + {M}_{i{\sigma _z}}^T{V_3}) + {c_2}({e^{ - ikx}}{M_{{\sigma _x}}}{V_5} + {e^{iky}}{M_{{\sigma _y}}}{V_6} + {e^{  ikz}}{M_{i{\sigma _z}}}{V_3})\\
			{I_8}& = ({l_g}{M_0} - 3({c_1} + {c_2}){I_{4 \times 4}}){V_8} + {c_1}({M}_{{\sigma _x}}^T{V_6} + {M}_{{\sigma _y}}^T{V_5} + {M}_{i{\sigma _z}}^T{V_4}) + {c_2}({e^{ikx}}{M_{{\sigma _x}}}{V_6} + {e^{-iky}}{M_{{\sigma _y}}}{V_5} + {e^{ ikz}}{M_{i{\sigma _z}}}{V_4})
		\end{aligned}
		\end{equation}
	\end{widetext}
	where $I_{i}=[I_{i,a}(\omega),I_{i,b}(\omega),I_{i,c}(\omega),I_{i,d}(\omega)]^T$ and $V_{i}=[V_{i,a}(\omega),V_{i,b}(\omega),V_{i,c}(\omega),V_{i,d}(\omega)]^T$ represent the current and voltage vectors at site $i$, respectively. The admittances of the circuit elements are given by $lg=\frac{1}{i\omega L_g}$, $c_1=i\omega C_1$, and $c_2=i\omega C_2$, corresponding to the grounding inductor $L_g$, the intra-cell capacitor $C_1$, and the inter-cell capacitor $C_2$, respectively. By applying the unitary transformation matrix $F$ defined in Eq. \ref{S2_eq}, we transform the equations into the pseudo-spin basis and obtain:
	
	\begin{widetext}
		\begin{equation*} 
			\begin{aligned}
			{{ I}_1'}=&  - \left[ {2{l_g}\left( {\left( {\begin{array}{*{20}{c}}
						0&0\\0&2
				\end{array}} \right) \oplus {\sigma _0}} \right) + 3({c_1} + {c_2})\left( {{\sigma _0} \oplus {\sigma _0}} \right)} \right]{{V}_1'} + {c_1}\left[ {\left( {{\sigma _0} \oplus {\sigma _x}} \right){{ V}_3'} + \left( {{\sigma _z} \oplus {\sigma _y}} \right){{ V}_4'} + \left( {{\sigma _z} \oplus  - i{\sigma _z}} \right){{ V}_5'}} \right] \\& + {c_2}\left[ {{e^{ikx}}\left( {{\sigma _0} \oplus {\sigma _x}} \right){{ V}_3'} + {e^{iky}}\left( {{\sigma _z} \oplus {\sigma _y}} \right){{ V}_4'} + {e^{-ikz}}\left( {{\sigma _z} \oplus i{\sigma _z}} \right){{ V}_5'}} \right]\\
		{{I}_2'} =&  - \left[ {2{l_g}\left( {\left( {\begin{array}{*{20}{c}}
						0&0\\0&2
				\end{array}} \right) \oplus {\sigma _0}} \right) + 3({c_1} + {c_2})\left( {{\sigma _0} \oplus {\sigma _0}} \right)} \right]{{ V}_2'} + {c_1}\left[ {\left( {{\sigma _0} \oplus {\sigma _x}} \right){{ V}_4'} + \left( {{\sigma _z} \oplus {\sigma _y}} \right){{ V}_3'} + \left( {{\sigma _z} \oplus  - i{\sigma _z}} \right){{ V}_6'}} \right] \\& + {c_2}\left[ {{e^{ - ikx}}\left( {{\sigma _0} \oplus {\sigma _x}} \right){{ V}_4'} + {e^{ - iky}}\left( {{\sigma _z} \oplus {\sigma _y}} \right){{ V}_3'} + {e^{-ikz}}\left( {{\sigma _z} \oplus i{\sigma _z}} \right){{ V}_6'}} \right]\\
			{{ I}_3'} =&  - \left[ {2{l_g}\left( {\left( {\begin{array}{*{20}{c}}
							0&0\\0&2
					\end{array}} \right) \oplus {\sigma _0}} \right) + 3({c_1} + {c_2})\left( {{\sigma _0} \oplus {\sigma _0}} \right)} \right]{{ V}_3'} + {c_1}\left[ {\left( {{\sigma _0} \oplus {\sigma _x}} \right){{ V}_1'} + \left( {{\sigma _z} \oplus {\sigma _y}} \right){{ V}_2'} + \left( {{\sigma _z} \oplus  - i{\sigma _z}} \right){{ V}_7'}} \right] \\& + {c_2}\left[ {{e^{ - ikx}}\left( {{\sigma _0} \oplus {\sigma _x}} \right){{ V}_1'} + {e^{iky}}\left( {{\sigma _z} \oplus {\sigma _y}} \right){{ V}_2'} + {e^{-ikz}}\left( {{\sigma _z} \oplus i{\sigma _z}} \right){{ V}_7'}} \right]\\
			{{ I}_4'}=&  - \left[ {2{l_g}\left( {\left( {\begin{array}{*{20}{c}}
							0&0\\0&2
					\end{array}} \right) \oplus {\sigma _0}} \right) + 3({c_1} + {c_2})\left( {{\sigma _0} \oplus {\sigma _0}} \right)} \right]{{ V}_4'} + {c_1}\left[ {\left( {{\sigma _0} \oplus {\sigma _x}} \right){{ V}_2'} + \left( {{\sigma _z} \oplus {\sigma _y}} \right){{ V}_1'} + \left( {{\sigma _z} \oplus  - i{\sigma _z}} \right){{ V}_8'}} \right] \\& + {c_2}\left[ {{e^{ikx}}\left( {{\sigma _0} \oplus {\sigma _x}} \right){{ V}_2'} + {e^{-iky}}\left( {{\sigma _z} \oplus {\sigma _y}} \right){{ V}_1'} + {e^{-ikz}}\left( {{\sigma _z} \oplus i{\sigma _z}} \right){{ V}_8'}} \right]\\
				{{ I}_5'}=&  - \left[ {2{l_g}\left( {\left( {\begin{array}{*{20}{c}}
								0&0\\0&2
						\end{array}} \right) \oplus {\sigma _0}} \right) + 3({c_1} + {c_2})\left( {{\sigma _0} \oplus {\sigma _0}} \right)} \right]{{ V}_5'} + {c_1}\left[ {\left( {{\sigma _0} \oplus {\sigma _x}} \right){{ V}_7'} + \left( {{\sigma _z} \oplus {\sigma _y}} \right){{ V}_8'} + \left( {{\sigma _z} \oplus i{\sigma _z}} \right){V_1}'} \right] \\& + {c_2}\left[ {{e^{ikx}}\left( {{\sigma _0} \oplus {\sigma _x}} \right){{ V}_7'} + {e^{iky}}\left( {{\sigma _z} \oplus {\sigma _y}} \right){{ V}_8'} + {e^{  ikz}}\left( {{\sigma _z} \oplus  - i{\sigma _z}} \right){{ V}_1'}} \right]\\
				{{ I}_6'}=&  - \left[ {2{l_g}\left( {\left( {\begin{array}{*{20}{c}}
								0&0\\0&2
						\end{array}} \right) \oplus {\sigma _0}} \right) + 3({c_1} + {c_2})\left( {{\sigma _0} \oplus {\sigma _0}} \right)} \right]{{ V}_6'} + {c_1}\left[ {\left( {{\sigma _0} \oplus {\sigma _x}} \right){{ V}_8'} + \left( {{\sigma _z} \oplus {\sigma _y}} \right){{ V}_7'} + \left( {{\sigma _z} \oplus i{\sigma _z}} \right){{ V}_2'}} \right] \\& + {c_2}\left[ {{e^{ - ikx}}\left( {{\sigma _0} \oplus {\sigma _x}} \right){{ V}_8'} + {e^{ - iky}}\left( {{\sigma _z} \oplus {\sigma _y}} \right){{ V}_7'} + {e^{  ikz}}\left( {{\sigma _z} \oplus  - i{\sigma _z}} \right){{ V}_2'}} \right]
					\end{aligned}
			\end{equation*}
		\end{widetext}
				
	\begin{widetext}
		\begin{equation} 
			\begin{aligned}
			{{ I}_7'}=&  - \left[ {2{l_g}\left( {\left( {\begin{array}{*{20}{c}}
							0&0\\0&2
					\end{array}} \right) \oplus {\sigma _0}} \right) + 3({c_1} + {c_2})\left( {{\sigma _0} \oplus {\sigma _0}} \right)} \right]{{ V}_7'} + {c_1}\left[ {\left( {{\sigma _0} \oplus {\sigma _x}} \right){{ V}_5'} + \left( {{\sigma _z} \oplus {\sigma _y}} \right){{ V}_6'} + \left( {{\sigma _z} \oplus i{\sigma _z}} \right){{ V}_3'}} \right]  \\&+ {c_2}\left[ {{e^{ - ikx}}\left( {{\sigma _0} \oplus {\sigma _x}} \right){{ V}_5'} + {e^{iky}}\left( {{\sigma _z} \oplus {\sigma _y}} \right){{ V}_6'} + {e^{  ikz}}\left( {{\sigma _z} \oplus  - i{\sigma _z}} \right){{ V}_3'}} \right]\\
			{{ I}_8'}=&  - \left[ {2{l_g}\left( {\left( {\begin{array}{*{20}{c}}
							0&0\\0&2
					\end{array}} \right) \oplus {\sigma _0}} \right) + 3({c_1} + {c_2})\left( {{\sigma _0} \oplus {\sigma _0}} \right)} \right]{{ V}_8'} + {c_1}\left[ {\left( {{\sigma _0} \oplus {\sigma _x}} \right){{ V}_6'} + \left( {{\sigma _z} \oplus {\sigma _y}} \right){{ V}_5'} + \left( {{\sigma _z} \oplus i{\sigma _z}} \right){{ V}_4'}} \right] \\&+ {c_2}\left[ {{e^{ikx}}\left( {{\sigma _0} \oplus {\sigma _x}} \right){{ V}_6'} + {e^{-iky}}\left( {{\sigma _z} \oplus {\sigma _y}} \right){{ V}_5'} + {e^{ ikz}}\left( {{\sigma _z} \oplus  - i{\sigma _z}} \right){{ V}_4'}} \right]
		\end{aligned}
		\end{equation}
	\end{widetext}

	where ${ I'(\omega)}_i=FI_i=[I_{i,0},I_{i,2},I_{i,\uparrow},I_{i,\downarrow}]^T$, ${ V'}_i=FV_i=[V_{i,0},V_{i,2},V_{i,\uparrow},V_{i,\downarrow}]^T$, $ \sigma_{x(y,z)}$ is Pauli matrix and $ \sigma_{0}$ is unity matrix. Consider the input current $I_i(\omega)=0$ and the pseudo-spin space $\tilde V_i(\omega)=[V_{i,\uparrow}(\omega),V_{i,\downarrow}(\omega)]^T$, kirchhoff's equctions are:
	\begin{widetext}
		\begin{equation} 
		\begin{aligned}
			\left[ {2{l_g} + 3({c_1} + {c_2})} \right]{\sigma _0}{{\tilde V}_1}& = \left( {{c_1} + {c_2}{e^{ikx}}} \right){\sigma _x}{{\tilde V}_3} + \left( {{c_1} + {c_2}{e^{iky}}} \right){\sigma _y}{{\tilde V}_4} - i\left( {{c_1} - {c_2}{e^{-ikz}}} \right){\sigma _z}{{\tilde V}_5}\\
			\left[ {2{l_g} + 3({c_1} + {c_2})} \right]{\sigma _0}{{\tilde V}_2}& = \left( {{c_1} + {c_2}{e^{ - ikx}}} \right){\sigma _x}{{\tilde V}_4} + \left( {{c_1} + {c_2}{e^{ - iky}}} \right){\sigma _y}{{\tilde V}_3} - i\left( {{c_1} - {c_2}{e^{-ikz}}} \right){\sigma _z}{{\tilde V}_6}\\
			\left[ {2{l_g} + 3({c_1} + {c_2})} \right]{\sigma _0}{{\tilde V}_3} &= \left( {{c_1} + {c_2}{e^{ - ikx}}} \right){\sigma _x}{{\tilde V}_1} + \left( {{c_1} + {c_2}{e^{iky}}} \right){\sigma _y}{{\tilde V}_2} - i\left( {{c_1} - {c_2}{e^{-ikz}}} \right){\sigma _z}{{\tilde V}_7}\\
			\left[ {2{l_g} + 3({c_1} + {c_2})} \right]{\sigma _0}{{\tilde V}_4}& = \left( {{c_1} + {c_2}{e^{ikx}}} \right){\sigma _x}{{\tilde V}_2} + \left( {{c_1} + {c_2}{e^{ - iky}}} \right){\sigma _y}{{\tilde V}_1} - i\left( {{c_1} - {c_2}{e^{-ikz}}} \right){\sigma _z}{{\tilde V}_8}\\
			\left[ {2{l_g} + 3({c_1} + {c_2})} \right]{\sigma _0}{{\tilde V}_5}& = \left( {{c_1} + {c_2}{e^{ikx}}} \right){\sigma _x}{{\tilde V}_7} + \left( {{c_1} + {c_2}{e^{iky}}} \right){\sigma _y}{{\tilde V}_8} + i\left( {{c_1} - {c_2}{e^{  ikz}}} \right){\sigma _z}{{\tilde V}_1}\\
			\left[ {2{l_g} + 3({c_1} + {c_2})} \right]{\sigma _0}{{\tilde V}_6}& = \left( {{c_1} + {c_2}{e^{ - ikx}}} \right){\sigma _x}{{\tilde V}_8} + \left( {{c_1} + {c_2}{e^{ - iky}}} \right){\sigma _y}{{\tilde V}_7} + i\left( {{c_1} - {c_2}{e^{  ikz}}} \right){\sigma _z}{{\tilde V}_2}\\
			\left[ {2{l_g} + 3({c_1} + {c_2})} \right]{\sigma _0}{{\tilde V}_7}& = \left( {{c_1} + {c_2}{e^{ - ikx}}} \right){\sigma _x}{{\tilde V}_5} + \left( {{c_1} + {c_2}{e^{iky}}} \right){\sigma _y}{{\tilde V}_6} + i\left( {{c_1} - {c_2}{e^{  ikz}}} \right){\sigma _z}{{\tilde V}_3}\\
			\left[ {2{l_g} + 3({c_1} + {c_2})} \right]{\sigma _0}{{\tilde V}_8} &= \left( {{c_1} + {c_2}{e^{ikx}}} \right){\sigma _x}{{\tilde V}_6} + \left( {{c_1} + {c_2}{e^{ - iky}}} \right){\sigma _y}{{\tilde V}_5} + i\left( {{c_1} - {c_2}{e^{ ikz}}} \right){\sigma _z}{{\tilde V}_4}
		\end{aligned}
		\end{equation}
	\end{widetext}
	
	The effective admittance matrix in pseudopsin space $\tilde J(\omega) \tilde V=0 $, and $\tilde J(\omega)/(i\omega)=H_{circuit}(k)-J_g(\omega)$ can give by:
	\begin{widetext}
		\begin{equation} 
			\begin{tiny}
			\begin{aligned}
				\hspace{-20mm}
				H_{circuit} =&\left( {\begin{array}{*{20}{c}}
						0&0&0&0&0&{{C_1} + {C_2}{e^{ikx}}}&0&{ - i\left( {{C_1} + {C_2}{e^{iky}}} \right)}\\
						0&0&0&0&{{C_1} + {C_2}{e^{ikx}}}&0&{i\left( {{C_1} + {C_2}{e^{iky}}} \right)}&0\\
						0&0&0&0&0&{ - i\left( {{C_1} + {C_2}{e^{ - iky}}} \right)}&0&{{C_1} + {C_2}{e^{ - ikx}}}\\
						0&0&0&0&{i\left( {{C_1} + {C_2}{e^{ - iky}}} \right)}&0&{{C_1} + {C_2}{e^{ - ikx}}}&0\\
						0&{{C_1} + {C_2}{e^{ - ikx}}}&0&{ - i\left( {{C_1} + {C_2}{e^{iky}}} \right)}&0&0&0&0\\
						{{C_1} + {C_2}{e^{ - ikx}}}&0&{i\left( {{C_1} + {C_2}{e^{iky}}} \right)}&0&0&0&0&0\\
						0&{ - i\left( {{C_1} + {C_2}{e^{ - iky}}} \right)}&0&{{C_1} + {C_2}{e^{ikx}}}&0&0&0&0\\
						{i\left( {{C_1} + {C_2}{e^{ - iky}}} \right)}&0&{{C_1} + {C_2}{e^{ikx}}}&0&0&0&0&0\\
						{i\left( {{C_1} - {C_2}{e^{  ikz}}} \right)}&0&0&0&0&0&0&0\\
						0&{ - i\left( {{C_1} - {C_2}{e^{  ikz}}} \right)}&0&0&0&0&0&0\\
						0&0&{i\left( {{C_1} - {C_2}{e^{  ikz}}} \right)}&0&0&0&0&0\\
						0&0&0&{ - i\left( {{C_1} - {C_2}{e^{  ikz}}} \right)}&0&0&0&0\\
						0&0&0&0&{i\left( {{C_1} - {C_2}{e^{  ikz}}} \right)}&0&0&0\\
						0&0&0&0&0&{ - i\left( {{C_1} - {C_2}{e^{  ikz}}} \right)}&0&0\\
						0&0&0&0&0&0&{ i\left( {{C_1} - {C_2}{e^{  ikz}}} \right)}&0\\
						0&0&0&0&0&0&0&{ - i\left( {{C_1} - {C_2}{e^{  ikz}}} \right)}
				\end{array}}   \bullet  \bullet  \bullet  \right.
				\\
				&\left. \bullet  \bullet  \bullet {\begin{array}{*{20}{c}}
						{ - i\left( {{C_1} - {C_2}{e^{-ikz}}} \right)}&0&0&0&0&0&0&0\\
						0&{i\left( {{C_1} - {C_2}{e^{-ikz}}} \right)}&0&0&0&0&0&0\\
						0&0&{ - i\left( {{C_1} - {C_2}{e^{-ikz}}} \right)}&0&0&0&0&0\\
						0&0&0&{i\left( {{C_1} - {C_2}{e^{-ikz}}} \right)}&0&0&0&0\\
						0&0&0&0&{ - i\left( {{C_1} - {C_2}{e^{-ikz}}} \right)}&0&0&0\\
						0&0&0&0&0&{i\left( {{C_1} - {C_2}{e^{-ikz}}} \right)}&0&0\\
						0&0&0&0&0&0&{ - i\left( {{C_1} - {C_2}{e^{-ikz}}} \right)}&0\\
						0&0&0&0&0&0&0&{i\left( {{C_1} - {C_2}{e^{-ikz}}} \right)}\\
						0&0&0&0&0&{{C_1} + {C_2}{e^{ikx}}}&0&{ - i\left( {{C_1} + {C_2}{e^{iky}}} \right)}\\
						0&0&0&0&{{C_1} + {C_2}{e^{ikx}}}&0&{i\left( {{C_1} + {C_2}{e^{iky}}} \right)}&0\\
						0&0&0&0&0&{ - i\left( {{C_1} + {C_2}{e^{ - iky}}} \right)}&0&{{C_1} + {C_2}{e^{ - ikx}}}\\
						0&0&0&0&{i\left( {{C_1} + {C_2}{e^{ - iky}}} \right)}&0&{{C_1} + {C_2}{e^{ - ikx}}}&0\\
						0&{{C_1} + {C_2}{e^{ - ikx}}}&0&{ - i\left( {{C_1} + {C_2}{e^{iky}}} \right)}&0&0&0&0\\
						{{C_1} + {C_2}{e^{ - ikx}}}&0&{i\left( {{C_1} + {C_2}{e^{iky}}} \right)}&0&0&0&0&0\\
						0&{ - i\left( {{C_1} + {C_2}{e^{ - iky}}} \right)}&0&{{C_1} + {C_2}{e^{ikx}}}&0&0&0&0\\
						{i\left( {{C_1} + {C_2}{e^{ - iky}}} \right)}&0&{{C_1} + {C_2}{e^{ikx}}}&0&0&0&0&0
				\end{array}} \right)
			\end{aligned}
		\end{tiny}
		\end{equation}
\end{widetext}
	
	We find that $H_{circuit}$  takes the same form as the model Hamiltonian $H$ presented in Eq. (2) of the main text. The eigenvalue is given by $ J_g=\left[3({C_1} + {C_2})-2/(\omega^2L_g)\right] I_{16\times 16}$, where 
	$I_{16\times 16}$ is the identity matrix. The most direct mathematical correspondence between the eigen-energies in the tight-binding model and the eigen-frequencies in the circuit metamaterials model is therefore: 
	\begin{equation}
		E = 3({C_1} + {C_2}) - 2/({\omega ^2}{L_g})\infty \frac{1}{{{\omega ^2}}}
	\end{equation}
	When the excitation frequency $ \omega $ satisfies $Det(\tilde J(\omega))=0$, the node voltages theoretically diverge, indicating that the circuit metamaterials reach a resonant state. In practice, due to parasitic effects—such as the inherent resistance in inductors—the node voltages do not become infinite delta functions but instead exhibit finite peaks with specific heights and widths at the resonance frequencies. By scanning the input frequency and measuring the voltage response at each node, one can identify these resonant frequencies. Subsequently, performing a Fourier transform on the spatial voltage distribution allows reconstruction of the dispersion relation in momentum space.
	
	\section*{APPENDIX C: SAMPLE FABRICATIONS AND MEASUREMENTs}
	We design the 2D printed circuit boards (PCBs) in the $x-y$ plane and realize the full 3D circuit metamaterials by stacking multiple 2D PCBs. The 2D PCBs are designed using Altium Designer (AD) software. Each carefully designed PCB consists of four layers of traces arranged to implement complex spin–orbit coupling (SOC) bridges along the  $\hat{x}$ and $\hat{y}$ directions. The SOC couplings between adjacent PCBs in the $\hat{z}$ direction are realized via four capacitors mounted on one PCB, which are then connected to the neighboring board through jumper wires. To minimize parasitic inductance effects in the experiment, all PCB traces maintain relatively large spacing of 0.5 mm. Four ground through-holes are placed at the corners of each PCB to facilitate stacking via copper pillar connections. These copper pillars not only ensure consistent grounding across all PCBs but also provide mechanical stability, supporting the entire 3D circuit metamaterials without bending under its own weight, while maintaining sufficient hollow space for excitation and measurement. Signal injection and detection are performed through BNC connectors soldered onto the PCBs. The imperfections in circuit components are controlled within approximately 5\%, ensuring faithful simulation of the target lattice model. Additionally, the boundary conditions are designed as follows:
	
	\begin{figure*}[htbp]
		\includegraphics[width=0.5\textwidth,height=0.35\textwidth]{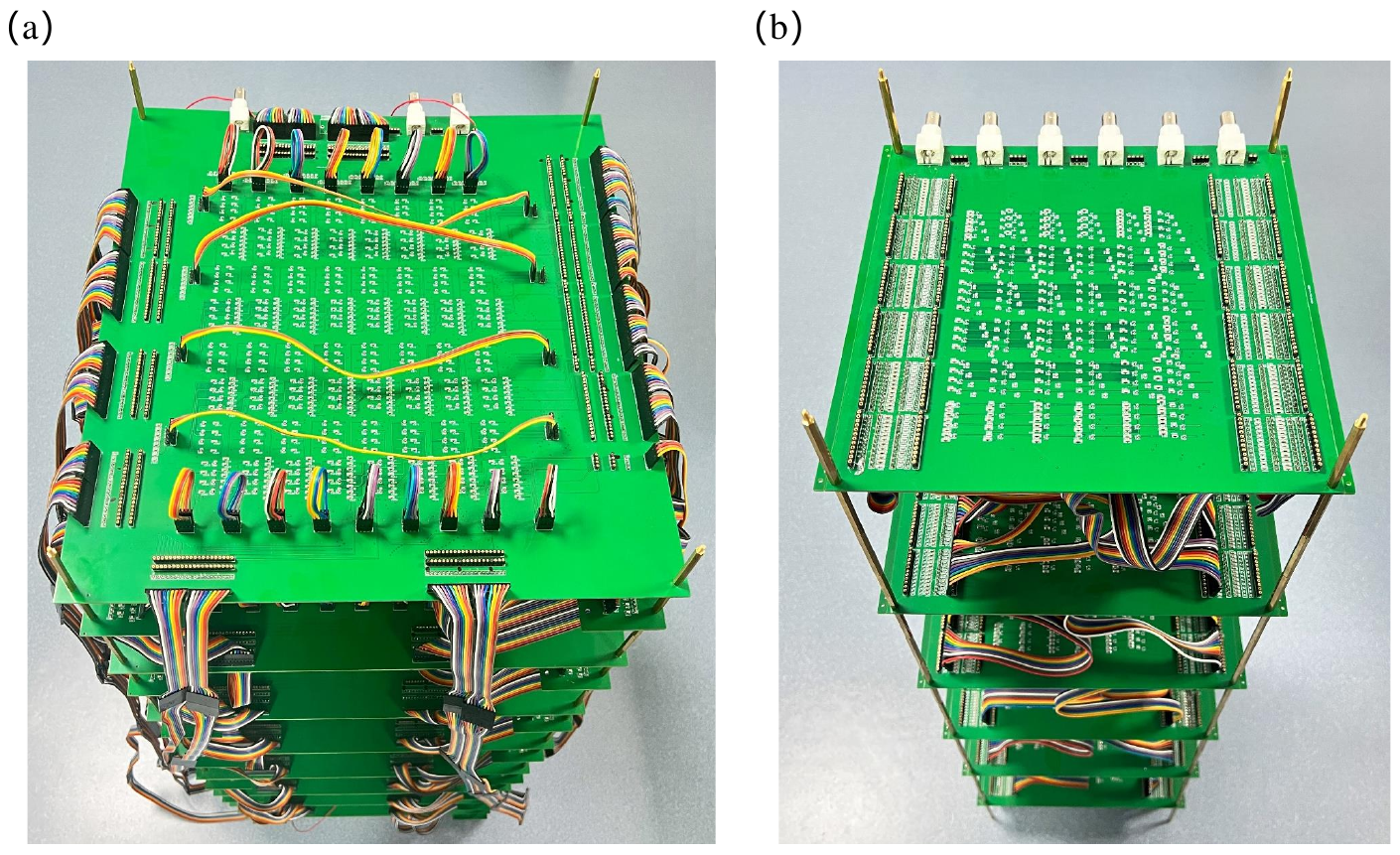}
		\caption{Experimental setup for 3D Weyl Semimetal circuit metamaterials (a) and 3D SOC HOTI circuit metamaterials (b).}
		\label{S3}
	\end{figure*}
	
	For 3D Weyl Semimetal circuit metamaterials: (i) Periodic boundary conditions are applied along the $ \hat{u}=\hat{x}+\hat{y} $ direction with traces on the PCB; (ii) Periodic boundary conditions are applied on the $\hat{z}$ direction using external wire connections; (iii) Open boundary conditions are applied along the $ \hat{v}=\hat{x}-\hat{y} $ direction.
	For 3D SOC HOTI circuit metamaterials: (i) Open boundary conditions are applied along the $\hat{x}$, $\hat{y}$, and $\hat{z}$ direction.
	
	The complete experimental setup is illustrated in Fig. \ref{S3}. The component parameters are selected as follows. For the 3D Weyl semimetal circuit metamaterials, inductors with $L_g=10$ $\mu$H and a tolerance of ±5\% are used. The capacitors $ C_{1} $ and $ C_{2} $ have identical specifications, each with a capacitance of 3.3 nF and ±5\% tolerance. For the 3D SOC third-order topological insulator circuit metamaterials, inductors of $L_g=3.3$ $\mu$H  with ±5\% tolerance are employed, together with capacitors $ C_1=330$ pF and $ C_2=3.3$ nF. The voltage at each node—including both amplitude and phase—is measured using a Keysight vector network analyzer (model E5061B).

	\section*{APPENDIX D: GROUNDING SETTING}
	\begin{figure*}[htbp]
		\includegraphics[width=0.55\textwidth,height=0.27\textwidth]{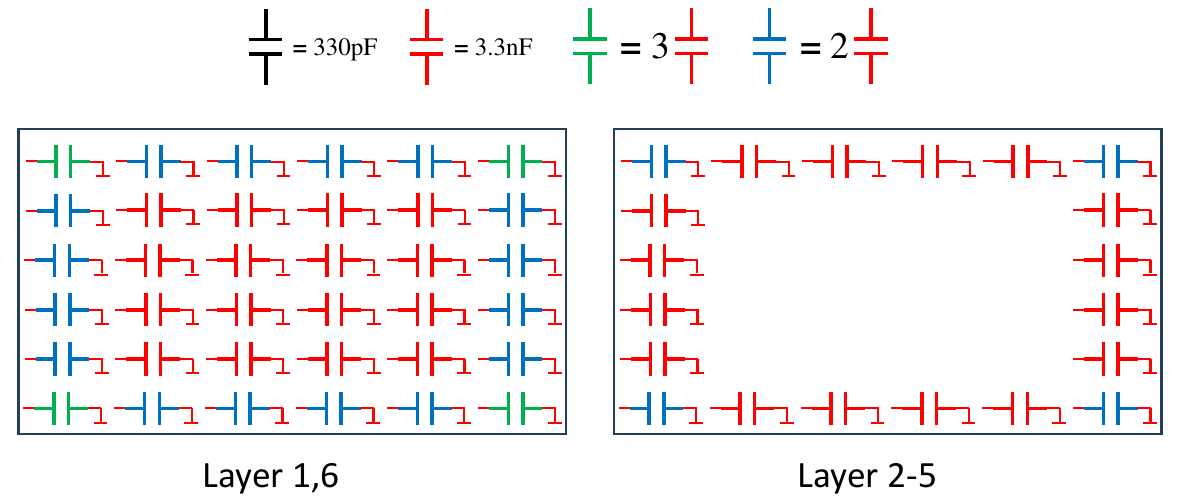}
		\caption{Experimental grounding configuration for 3D SOC HOTI circuit metamaterials.}
		\label{S4}
	\end{figure*}
	One important aspect to note regarding open boundary conditions is the presence of diagonal terms in $J_g$ arising from couplings in the circuit Hamiltonian. When switching to open boundary conditions, an additional coupling must be introduced to compensate for the reduction in $J_g$. Consequently, the grounding conditions for circuit sites located in the 3D bulk, 2D surfaces, 1D hinges, and 0D corners differ and must be carefully engineered. This difference stems from the varying number of connected capacitors at these nodes. The experimental grounding configuration for our 3D SOC HOTI circuit metamaterial is detailed in Fig. \ref{S4}.
	
	\section*{APPENDIX E: LOCAL DENSITY OF STATES MEASUREMENTS}
	\setcounter{equation}{0}
	\renewcommand{\theequation}{E\arabic{equation}}
	In circuit systems, the LDOS can be experimentally obtained by measuring the real part of the electrical impedance. The LDOS is defined as:
	\begin{equation}
		\rho(\varepsilon,x)=\sum_{n}{|\varphi _n^x|^2\delta(\varepsilon- \varepsilon_n)},
	\end{equation}
	where $\varphi_n^x$ denotes the component of the eigenstate at lattice site $x$, $\varepsilon_n$ is the corresponding eigenenergy, and $\delta(\varepsilon - \varepsilon_n)$ is the Dirac delta function. To facilitate numerical and experimental processing, the delta function is often approximated by a Lorentzian function:
	\begin{equation}
		\delta(x)=\frac{1}{\pi}\lim_{\varepsilon \to 0}\frac{\varepsilon}{x^2+\varepsilon^2}
	\end{equation}
	Accordingly, the LDOS can be rewritten as:
	\begin{equation}
		\rho(\varepsilon,x)=\frac{1}{\pi}\lim_{\varepsilon \to 0}\sum_{n}{\frac{\varepsilon}{x^2+\varepsilon^2}|\varphi_n^x|^2},	
	\end{equation}
	In the circuit network, the admittance matrix is defined as $J(\omega) = i\omega [H - E(\omega)]$, and its inverse is the Green's function:
	$G(\omega)= J(\omega)^{-1}=\frac{1}{i\omega}\sum_{n}{\left| {{\psi _n}} \right\rangle \frac{1}{E_n - E(\omega)}\left\langle {{\psi _n}} \right|}$, where $E_n$ represents the eigenvalue of the Hamiltonian $H$. The node voltages relate to the injected currents by $\mathbf{V}(\omega) = G(\omega) \mathbf{I}(\omega)$. Under experimental conditions, current is injected only at the $a$-th node, i.e., $\mathbf{I} = (0, \dots, I_a, \dots, 0)^\text{T}$, so the voltage at this node is:
	\begin{equation}
		V_a(\omega)=G_{aa}(\omega) I_a(\omega),
	\end{equation}
	where $V_a(\omega)$ is the $a$-th component of $ \textbf{V}(\omega) $ and $G_{aa}(\omega)$ is the $a$-th diagonal element of the Green's function $G(\omega)$. The impedance between node $a$ and the ground is given by:
	\begin{equation}
		Z_a(\omega)=\frac{V_n(\omega)}{I_n(\omega)}=G_{aa}(\omega),
	\end{equation}
	Expanding this expression and introducing a convergence factor $\varepsilon$, we obtain:
	\begin{equation}
		\begin{aligned}
			G_{aa}(\omega)	&=\lim_{\varepsilon \to 0} \sum_{n}{\frac{|\psi_{n,a}|^2}{i\omega\left(E_n-E(\omega)\right)+\varepsilon}}\\&=\lim_{\varepsilon \to 0} \sum_{n}{\frac{[-i\omega\left(E_n-E(\omega)\right)+\varepsilon]|\psi_{n,a}|^2}{[\omega\left(E_n-E(\omega)\right)]^2+\varepsilon^2}}
		\end{aligned}
	\end{equation}
	Extracting the real part yields:
	\begin{equation}
		\text{Re}[Z_a(\omega)]=\lim_{\varepsilon \to 0} \sum_n{\frac{\varepsilon|\psi_{n,a}|^2}{\left[\omega \left(E_n-E(\omega)\right)\right]^2+\varepsilon^2}}
	\end{equation}
	Using the Lorentzian representation of the delta function, this can be expressed as:
	\begin{equation}
		\text{Re}[Z_a(\omega)] = \pi \sum_n \delta[\omega(E_n - E(\omega))] |\psi_{n,a}|^2.
	\end{equation}
	Further, we derive:
	\begin{equation}
		\text{Re}[Z_a(f)] = \frac{1}{2f} \sum_n \delta(E_n - E(f)) |\psi_{n,a}|^2.
	\end{equation}
	Then, the LDOS can be extracted from the experimental data using the formula:
	\begin{equation}
		\rho (f,a)=2f\text{Re}[Z_a(f)]
	\end{equation}
	This expression provides a practical basis for experimentally determining the LDOS.
	
	\section*{APPENDIX F: REAL-SPACE TOPOLOGICAL INVARIANTeal-space topological invariant}
	\setcounter{equation}{0}
	\renewcommand{\theequation}{F\arabic{equation}}
	Consider a chiral-symmetric Hamiltonian ${H}$, which satisft $ \Gamma {H}\Gamma^{-1} =-{H}$, where $ \Gamma$ is the chiral operator. In the basis in which the chiral operator is $\Gamma=\sigma_z$, the Hamiltonian ${H}$ take the form:
	\begin{equation}
		{H} = \left( {\begin{array}{*{20}{c}}
				0&h\\
				{h^\dag }&0
		\end{array}} \right)
	\end{equation}
	which allow a partition of the lattice into two sublattices, A and B, with opposite chiral charge. The multipole chiral number (MCN) for third-order topological phases with chiral symmetry are based on extensions  of 1D real space index. Consider a lattice in 3D with $L_j$ unit cells along direction $ j $ ($ j=x,y,z $). Each unit cell is labeled by $\textbf{R}=(x, y, z)$.
	we define the sublattice multipole moment operators:
	\begin{equation}
		{O^S}_{xyz} = \sum\limits_{{\bf{R}},\alpha  \in S} {\left| {{\bf{R}},\alpha } \right\rangle \text{Exp}( - i\frac{{2\pi xyz}}{{{L_x}{L_y}{L_z}}})\left\langle {{\bf{R}},\alpha } \right|} 	
	\end{equation}
	where $ S=A,B$ and $\left| {{\bf{R}},\alpha } \right\rangle ={c^\dag }_{R,\alpha }\left| 0 \right\rangle $, ${c^\dag }_{R,\alpha }$ creates an electron at orbital $\alpha$ of unit cell $R$. The integer invariants for chiral-symmetric third-order topological phases in 3D is:
	\begin{equation}
		{N_{xyz}} = \frac{1}{{2\pi i}}Tr\log ({{\bar O}^A}_{xyz}{{\bar O}^{B\dag }}_{xyz})
	\end{equation}
	where ${{\bar O}^S}_{xyz} = U_S^\dag {O^S}_{xyz}{U_S}$ is  sublattice multipole moment operators projected into the space $ U_S$. $ U_S$ give by singular value decomposition (SVD) of $ h=U_A \Sigma U^\dag_B $.

	\end{document}